\documentclass[10pt]{article}

\usepackage{times}

\usepackage{color} 
\usepackage{lmodern}
\usepackage{multicol}
\usepackage{subfigure}
\usepackage{graphicx} 
\usepackage{booktabs}
\usepackage{bbm}
\usepackage{bm}
\usepackage{amsmath}
\usepackage{amsfonts}
\usepackage{mathtools}
\usepackage{booktabs}
\usepackage{tikz}
\usepackage{nicefrac}
\usetikzlibrary{positioning}
\usetikzlibrary{arrows}
\usepackage[round]{natbib}
\usepackage{algorithm}
\usepackage{algorithmic}

\usepackage{fancybox} 
\usepackage{pgfplots}

\DeclareMathOperator*{\argmin}{argmin}
\DeclareMathOperator*{\argzero}{argzero}

\providecommand{\keywords}[1]
{
  \large	
  \textbf{\textit{Keywords---}} #1
}

\usepackage[utf8]{inputenc}
\usepackage[english]{babel}
\usepackage{amsthm}
\usepackage{enumitem}

\usepackage[pdfpagelabels, hyperindex=true,pdftitle={},pdfauthor={St\'ephane Guerrier},colorlinks=TRUE, pagebackref=false,citecolor=blue, urlcolor = blue, plainpages=false, pdfpagelabels]{hyperref} 

\newtheorem{definition}{Definition}
\newtheorem{remark}{Remark}
\newtheorem{prop}{Proposition}
\newtheorem{corol}{Corollary}

\newcounter{lastnote}

\title{Perturbed M-Estimation: A Further Investigation of\\ Robust Statistics for Differential Privacy} 

\author
{Aleksandra Slavkovic,$^{1}$ Roberto Molinari$^{2}$\footnote{This work was in part done when Molinari was a Lindsay Visiting Assistant Professor at Penn State University} \\
\\
\normalsize{$^{1}$Department of Statistics, Penn State University} \\
\normalsize{$^{2}$Department of Mathematics and Statistics, Auburn University}
}

\date{}

\begin{document} 


\baselineskip24pt


\maketitle


\begin{abstract}
Differential Privacy (DP) provides an elegant mathematical framework for defining a provable disclosure risk in the presence of arbitrary adversaries; it guarantees that whether an individual is in a database or not, the results of a DP procedure should be similar in terms of their probability distribution.While DP mechanisms are provably effective in protecting privacy, they often negatively impact the utility of the query responses, statistics and/or analyses that come as outputs from these  mechanisms. To address this problem, we use ideas from the area of robust statistics which aims at reducing the influence of outlying observations on statistical inference. Based on the preliminary known links between differential privacy and robust statistics, we modify the objective perturbation mechanism by making use of a new bounded function and define a bounded M-Estimator with adequate statistical properties. The resulting privacy mechanism, named ``Perturbed M-Estimation'', shows important potential in terms of improved statistical utility of its outputs as suggested by some preliminary results. These results consequently support the need to further investigate the use of robust statistical tools for differential privacy.
\end{abstract}

\keywords{Differential Privacy, Robust Statistics, Objective Perturbation, Utility, Parametric Inference, Hyperbolic Tangent/Cosine Function}

\section{Introduction} 
We live in a world of continuous data collection, storage, and sharing, with much of those data being sensitive, making data privacy a highly relevant societal topic\footnote{ See, for example, \url{https://www.nytimes.com/interactive/2019/opinion/internet-privacy-project.html}}. Steve Fienberg has recognized the importance of data privacy and confidentiality, and crucially the role that statistical science must play in this context. He had argued that the right methodology for collecting and sharing of sensitive data should rely on statistical principles of sampling, estimation and modeling, transparency of masking procedure, and the dualities of the data utility and the disclosure risk. Steve argued for these guiding principles in many congressional and government testimonies, and followed them in numerous scholarly contributions on the topic of data privacy and confidentiality. In his first technical contribution in this area, he proposed a
bootstrap-like approach for creating synthetic data, similar to the current synthetic data
methodology that relies on multiple imputation (\cite{fienberg1994radical}). Here we highlight a few additional representative publications of his --- for example, see \cite{fienberg1998disclosure} on perturbation of categorical data; 
\cite{duncan2001disclosure} for general disclosure principles and links to information loss; 
\cite{trottini2002modelling} on Bayesian modeling of disclosure risk; \cite{fienberg2005preserving} on links between privacy-preserving data mining and contingency table releases; 
\cite{fienberg2008valid} on distributed regression analysis and secure multi party computation; \cite{fienberg2010differential} on data privacy links to algebraic statistics and log-linear models; 
\cite{hall2011secure} on how to perform distributed regression using homomorphic encryption; 
\cite{wang2016average} on KL-privacy and its links to differential privacy; and 
\cite{lei2018differentially}) on model selection under differential privacy; for a more comprehensive list see \cite{Slavkovic_Vilhuber_2018}.

Statistical data privacy, traditionally referred to as statistical disclsoure limitation or control (SDL or SDC), is the branch of statistics concerned with limiting identifying information in released data and summaries while maintaining their utility for valid statistical inference. It has a rich history for both methodological developments and applications for “safe” release of altered (or masked) microdata and tabular data (see \cite{dalenius1977privacy}, \cite{willenborg1996statistical}, \cite{fienberg2011}, \cite{hundepool}, and references therein). Besides traditional methods such as supression and aggregation, many modern methods rely on sampling and modeling, such as synthetic data (e.g., \cite{rubin1993statistical}, \cite{reiter2005using}, \cite{snoke2018}), and aim to frame data privacy as a statistical problem that requires treating both the data utility and the disclosure risk as random variables. However, they often fall short of allowing for the transparency of masking procedures, which is important in order to achieve the right statistical inference, not the individual identification.  Furthermore, the onslaught of big data has presented new challenges for traditional statistical data privacy methodology and  the so-called ``reconstruction theorem" (e.g., see \cite{dinur2003revealing} and \cite{Garfinkel2018}) has identified a flaw in a probabilistic notion of disclosure as proposed by \cite{dalenius1977privacy}. Many practical examples have demonstrated increased privacy risk from the released data or summaries in presence of other `axuilliary' data that were previously either not considered or simply were not as readily accessible; see \cite{dwork2017exposed} for a survey of such attacks, and recent claims related to issues with the U.S. Census data (\cite{abowd2018staring}).

Differential Privacy (DP) has emerged from theoretical computer science with a goal of designing transparent privacy mechanisms/methods with mathematically provable disclosure risk in the presence of adversaries with arbitrary priors, unlimited side information, and unbounded computational
power, e.g., see \cite{dwork2006calibrating} for the original proposal and \cite{slavkovic-steve-dp} and \cite{Slavkovic_Vilhuber_2018} for Steve's role in bringing computer scientists, statisticians and practitioners together to forge the new directions of formal privacy.  Differential privacy guarantees that whether an individual is in a database or not, the
results of a DP method should be similar in terms of their probability distribution;  this limits the ability of an adversary to
infer about any particular individual (unit) in the database and at the same time allows the data analyst to carry out inference on a distribution not sensitive to outliers.  DP quantifies the so-called privacy-loss budget, $\epsilon$, to how much the answer to a question or statistic is changed given the absence or presence of the most extreme possible person in the population.

Understanding the above risk-utility tradoffs under formal privacy constraints such as those imposed by DP and linking them to fundamental statistical concepts has been one of the key recent research threads in data privacy, as there are serious implications on how we carry valid statistical inferences if data are to be shared under the DP framework. \cite{wasserman2010statistical} were among the first to underline these links focusing on density estimation and offering a statistically-flavored interpretation of DP. Over the past decade numerous works have explored these links in different settings including parameter estimation (\cite{smith2011privacy}, \cite{duchi2013local}), hypothesis testing (\cite{vu2009differential}, \cite{wang2015revisiting}, \cite{gaboardi2016differentially}, \cite{awan2018differentially}, \cite{canonne2019structure}), confidence intervals (\cite{karwa2017finite}), model selection (\cite{lei2018differentially}), principal component analysis (\cite{chaudhuri2013near}, \cite{awan2019benefits}), network data (\cite{karwa2016inference}), and functional data analysis (\cite{hall2013differential}, \cite{mirshani2019formal}), to name a few. 

 \cite{dwork2009differential} were the first to investigate links between differential privacy and robust statistics (e.g., see~\cite{huber2011robust}).  One of the fundamental concepts behind differential privacy is to define the maximum amount of change a query or statistic can undergo (\textit{sensitivity}) when one row in the database is added or replaced by another arbitrary row. Once this sensitivity is defined, differentially private mechanisms add a proportional amount of noise in order to hide whether a change in output is due to a change in row or to the added noise; the amount of noise grows with the sensitivity of the query/statistic. Robust statistics aims at limiting the impact that an extreme observation can have on statistical estimation and inference. In this sense, using robust statistics can deliver statistics and/or analyses with bounded sensitivity. Based on this property, robust statistics can bound the (DP) sensitivity and therefore reduce the amount of noise required to ensure privacy and consequently improve utility of the private outputs. \cite{dwork2009differential} explore these links and make use of robust estimators (e.g., median and interquartile range) as a starting point for releasing differentially private estimators based on a Propose-Test-Release algorithm for interactive queries, while \cite{lei2011differentially} proposes the use of (bounded) M-Estimators applied to differentially private perturbed histograms in order to enhance the utility of statistical estimations under DP. \cite{chaudhuri2012convergence} study convergence rates of differentially private approximations to statistical estimators and propose the use of (bounded) M-Estimation within the exponential mechanism. Most recently,  \cite{avella2019privacy} proposed a statistical inference framework where noise is added to the M-Estimators in order to ensure privacy. 
 
 In this paper we investigate the use of functions with bounded derivatives, such as those used for M-Estimation in robust statistics, within the Objective Perturbation Mechanism (OPM), originally proposed in \cite{chaudhuri2011differentially}, and modified by \cite{kifer2012private}. We propose a new convex and bounded function called the Robust Hyperbolic Tangent (RobHyt) function which can be used to produce a bounded M-Estimator with adequate statistical properties which itself can be easily integrated within the OPM framework. More specifically, we study the statistical consistency of this bounded non-private M-Estimator. In the non-private setting, the choice of the bounding parameter (that is of the tuning constant) is usually made based on the asymptotic properties of the non-private M-Estimator. However, when integrating the M-Estimator in the OPM, the tuning constant can be used as a parameter regulating the trade-off between statistical efficiency and the amount of noise added for privacy. Thus, the non-private statistical properties of the proposed M-Estimator can provide a first rule to define this tuning constant when employed within the OPM. Based on the preliminary results, the resulting privacy mechanism, that we name the ``Perturbed M-Estimation'' mechanism, can greatly improve the utility of differentially private outputs while preserving the same level of privacy.

This paper is organized as follows. In Section \ref{sec.rob_dp_link} we provide a summary overview of important definitions for differential privacy and then make links between these definitions and the framework of robust statistics. In Section \ref{sec.pert_mest} we briefly introduce M-Estimation theory and propose the RobHyt function to deliver a bounded M-Estimator. This estimator is then used to build Perturbed M-Estimation by integrating it within the OPM. In Section \ref{sec.app_sim} we study the performance of the proposed method by using both the simulated and real-data examples, particularly focusing on gains in statistical utility in comparison to some existing methods. Finally, Section \ref{sec.conclusion} concludes and provides possible future avenues of research in the proposed direction.

\section{A Robust Parametric View of Differential Privacy}
\label{sec.rob_dp_link}

The basic idea behind differential privacy is to protect the privacy of an individual in the worst case scenario where an adversary is in possession of the data of all the other individuals in a database except for those of this particular individual. The release of differentially private data or analyses requires mechanisms (methods) to add noise, directly or via sampling, in such a way that an output of these mechanisms is (nearly) equally likely to occur whether or not an individual is included in a database. More formally, a mechanism $\mathcal{M}(\cdot)$ is defined to be $(\epsilon, \delta)$-differentially private if it respects the following condition
$$\mathbb{P}[\mathcal{M}(D) \in S] \leq e^{\epsilon}\mathbb{P}[\mathcal{M}(D') \in S] + \delta, $$
where $D$ and $D'$ are two databases that differ in one row (i.e., neighbouring databases) and $S$ is a set of outputs belonging to the range of $\mathcal{M}(\cdot)$. This definition implies that, for the same output $S$, the probability of observing it given the database $D'$ is within an ``$\epsilon$-range'' of the probability of observing the same output given the database $D$ plus an exception $\delta$. This must hold for all measurable sets $S$ and all pairs of databases $D$ and $D'$ that differ in one entry. The quantities $\epsilon$ and $\delta$ should be small. When $\delta = 0$ then we have so-called ``pure'' differential privacy, while the presence of a small $\delta$ (e.g., decreasing polynomially with the sample size $n$) allows for the data of (some) individuals to be released entirely with low probability $\delta$. The value $\epsilon $ is the  privacy  parameter,  or  the  privacy-loss  budget.   Smaller values correspond to more privacy, but as it approaches infinity there is no privacy guarantee.

DP mechanisms most often introduce some form of noise in the analysis (or data) or distort the problem definition underlying a query or estimation procedure in order to cover the variation due to the change in one individual's data. To determine the degree of ``variation'', and thus the amount of noise to be added, different notions of \textit{sensitivity} have been proposed and discussed in the privacy literature. The \textit{global sensitivity} is defined as
$$GS_f = \underset{D, D'}{\max} \|f(D) - f(D')\|,$$
where $f(\cdot)$ is any function (query, estimator, etc.) and this measure captures the maximum extent to which the function $f(\cdot)$ can vary between all possible combinations of neighboring databases. The \textit{local sensitivity}, 
$$LS_f = \underset{D'}{\max} \|f(D) - f(D')\| ,$$
 fixes the database of reference $D$ and determines the maximum variation considering all other possible neighboring databases $D'$. Other notions of sensitivity exist and other norms to determine them are also considered (see, for example, \cite{dwork2014algorithmic} and \cite{awan2018structure}, and references therein). These quantities are important for improving the risk-utility trade-offs. The smaller the sensitivity, the smaller the amount of noise is required for privacy, which typically leads to better \textit{utility} of the outputs, and possibly better management of the privacy-loss budget. In \cite{awan2018structure} and in this paper, we show that for the same privacy cost, we gain better utility and more usefulness of data if we propose ways of adjusting the sensitivities of the outputs.

The above notions of sensitivity which measure possible variations of estimating functions is strongly related to the notions underlying the framework of robust statistics. The next section highlights the similarities between these notions and justifies the investigation of robust statistical tools for the purposes of achieving differential privacy since, by reducing the sensitivity of estimators (functions $f(\cdot)$), robust statistical approaches can require less noise in order to deliver more useful differentially private outputs.

\subsection{Links with Robust Statistics}

As highlighted above, the notion of differential privacy and the concepts based on which differentially private mechanisms are proposed are intrinsically linked with notions of function (query) sensitivity, centered around the space of neighboring datasets. Robust statistics, on the other hand, focuses on the sensitivity of the function with respect to the quantity it is meant to compute (estimate) which, in general, corresponds to the output that would be observed if the function were applied to the entire population of reference. This is formalized within robust statistics by using a parametric framework where the population is described by an assumed parametric model $F_{\bm{\theta}}$, with $\bm{\theta} \in \Theta \subset \mathbb{R}^p$ being the parameter vector defining the model. The goal in this setting is to estimate the parameter vector $\bm{\theta}$ (e.g., the regression coefficients and residual variance) through an estimator (function) with appropriate statistical properties. 

However, the framework of robust statistics postulates that although we assume a model $F_{\bm{\theta}}$ for our data, this is at best an approximation to reality and what we actually observe is
\begin{equation}F_{\lambda} = (1 - \lambda) F_{\bm{\theta}} + \lambda G ,\label{eq:contamination-model}\end{equation} 
for small $\lambda > 0$ and with $G$ being an unspecified ``contamination'' model (see e.g., \cite{huber2011robust}, \cite{hampel1986robust}, \cite{maronna2019robust}). In this paradigm the goal of an estimation and optimization problem would be to recover the value of $\bm{\theta}$ as best as possible by reducing the impact of the unknown model $G$. More specifically, let us define an estimator as a functional $T(F)$ where $F$ is a general notation for a model (e.g., empirical or parametric). When we apply this functional to $F_{\lambda}$ we would want to obtain a good estimate that is output for the true value of $\bm{\theta}$, but this will depend on the properties of the functional. In order to determine these properties when observing $F_{\lambda}$, the notion of Influence Function (IF) was introduced (see \cite{hampel1974influence}) and is defined as follows
$$\text{IF}_T(z_0, F_{\bm{\theta}}) = \underset{\lambda \downarrow 0}{\lim} \frac{T((1 - \lambda)F_{\bm{\theta}} + \lambda \Delta_{z_0}) - T(F_{\bm{\theta}})}{\lambda} ,$$
where $\Delta_{z_0}$ is a point-mass distribution in an arbitrary point $z_0$ which plays the role of the model $G$. In general terms, this quantity can be interpreted as the impact that an infinitesimal amount of contamination can have on a given functional $T$.

The IF is therefore an important notion in robust statistics since it can be used as a measure to understand the possible extent of asymptotic bias with respect to $\bm{\theta}$ introduced by the presence of $G$. An additional measure that is based on the IF is given by the Gross Error Sensitivity (GES) defined as
$$\gamma(T, F_{\bm{\theta}}) = \underset{z_0}{\sup} \left| \text{IF}_T(z_0, F_{\bm{\theta}}) \right| .$$
The GES measures the maximum impact that any point-mass distribution $\Delta_{z_0}$ can have on the estimator $T$. Then an estimator $T$ is defined as being (B-)robust if the GES is bounded, that is if the IF is bounded --- which is a sufficient condition.

Taking a deeper look at these definitions one can see the similarities with the sensitivity definitions used for differential privacy.  Let $\lambda=\frac{1}{n}$ and assume that the empirical distribution $F_n$ (an estimator of $F_{\bm{\theta}}$) fully characterizes the database $D$. Then we could
reformulate the contamination model from equation (\ref{eq:contamination-model}) as$$F_{\lambda} = \frac{n - 1}{n}F_n + \frac{1}{n}z_0, $$
which resembles another definition in robust statistics, i.e., {\em the sensitivity curve}. These types of robust measures resemble the definition of \textit{local sensitivity} since they would measure the impact of one observation, $z_0$, on the database (model) of reference $D$ (i.e., $F_{\bm{\theta}}$). The notion of {\em global sensitivity}, on the other hand, would require a contamination model where all possible versions of $F_{\bm{\theta}}$ are considered.

As mentioned earlier, other works have explored the similarities of differential privacy notions with those of robust statistics highlighted above. For example, after defining the above robustness measures, \cite{chaudhuri2012convergence} use the notion of GES to deliver convergence rates for differentially private statistical estimators while \cite{avella2019privacy} uses this measure to calibrate the additive noise to deliver differentially private M-Estimators. In the next sections we explore another approach, suggested but not studied in \cite{chaudhuri2012convergence} and \cite{avella2019privacy}, where we investigate the use of {\em bounded} M-Estimation for differentially private estimation and prediction using the OPM. While empirical risk minimization, that objective perturbation is built on, can be classified as  M-Estimation, it is not straightforward to integrate the standard bounded functions for M-estimation as is. To address this problem, we propose a modified OPM that we call the Perturbed M-Estimation mechanism. More specifically, we propose the use of a new convex objective function, {\em RobHyt}, defining a bounded M-Estimator for which we first study its non-private statistical properties and convergence rates which then lead to its integration in a differentially private setting. 

\section{Perturbed M-Estimation}
\label{sec.pert_mest}
In this section we present the Perturbed M-Estimation mechanism designed by integrating a new bounded function into the OPM of \cite{kifer2012private}, thereby improving the overall utility of the differentially private output. Recall that the goal of robust statistics is to bound the impact of outlying observations on the output of an analysis. A popular class of estimators for this purpose is that of M-Estimators defined as
\begin{equation}
\label{def_mest}
    \bar{\bm{\theta}} = \underset{\bm{\theta} \in \bm{\Theta}}{\argmin} \frac{1}{n}\sum_{i = 1}^n \rho(\bm{\theta}; d_i),
\end{equation}
where $\bm{\theta}$ is a parameter of interest we aim to release, $\rho(\cdot)$ is a convex loss function and $d_i \in D \in \mathcal{D}^n$ is the $i^{th}$ row of a database with independent rows. In this form, the class of M-Estimators corresponds to the notion of empirical risk minimization. However, in order for the resulting estimator $\bar{\bm{\theta}}$ to be robust we require the derivative of the loss function to be bounded. The IF of an M-Estimator is given by
$$\text{IF}_T(z_0, F_{\bm{\theta}}) = - \psi(z_0, T(F_{\bm{\theta}}))\,B(T(F_{\bm{\theta}}), \psi)^{-1} ,$$
where $z_0 \in \mathbb{R}$ is an arbitrary point, $\psi(z) = \nicefrac{\partial}{\partial z} \, \rho(z)$ and $B(T(F_{\bm{\theta}}), \psi) = \nicefrac{\partial}{\partial \bm{\theta}} \, \mathbb{E}[\psi(\bm{\theta}; z)]$ (see \cite{hampel1986robust}). Thus, the IF of an M-Estimator is bounded if the $\psi$-function is bounded which justifies why, in many cases within the robust literature, M-Estimators are also expressed directly with respect to their derivative as follows
\begin{equation*}
    \bar{\bm{\theta}} = \underset{\bm{\theta} \in \bm{\Theta}}{\argzero} \frac{1}{n}\sum_{i = 1}^n \psi(\bm{\theta}; d_i),
\end{equation*}
which allows for reformulating the optimization problem in the form a system of estimating equations. The class of Maximum-Likelihood Estimators (MLE) can be represented as M-Estimators where $\rho(\cdot)$ would correspond to the negative log-likelihood and $\psi(\cdot)$ its derivative. But the MLE is not robust since, in general, the corresponding $\psi$-function is unbounded with respect to the data. Different functions have been proposed for $\rho(\cdot)$ in order to bound $\psi(\cdot)$, such as the Huber and Tukey Biweight functions (see e.g. \cite{hampel1981change}, \cite{maronna2019robust}). These functions, along with other bounded functions commonly used for robustness purposes, implicitly or explicitly assign weights to the residuals or score functions defined by the minimization problem thereby downweighing observations that lie far from the ``center'' of the assumed distribution of the residuals $F$. However, these functions typically have symmetric weights and can therefore be asymptotically biased with respect to the distribution of the residuals $F$ (for example if the latter is asymmetric). Hence, a correction factor is often added for Fisher consistency (e.g., see \cite{huber2011robust} and  \cite{cantoni2001robust}) which depends on the model $F$ and the chosen bounded function $\psi(\cdot)$.

The definition of an estimator as an M-Estimator has additional advantages from a point of view of \textit{parametric} statistical inference. Under a set of regularity conditions on the properties of the $\psi(\cdot)$ function and the parameter space, the asymptotic distribution of M-Estimators (see \cite{mises1947asymptotic}, \cite{hampel1986robust}) is 
\begin{equation*}
	\sqrt{n}(\bar{\bm{\theta}} - \bm{\theta}_0) \rightarrow \mathcal{N}\left(\bm{0}, \bm{\Sigma} \right),
\end{equation*}
where $\bm{\theta}_0$ represents the true parameter vector we aim to estimate and 
$$\bm{\Sigma} = \bm{M}_{\psi}(\bm{\theta}_0) \bm{Q}_{\psi}(\bm{\theta}_0) \bm{M}_{\psi}(\bm{\theta}_0)^T\,,$$
is the asymptotic covariance matrix where
$$\bm{M}_{\psi}(\bm{\theta}_0) = \frac{\partial}{\partial \bm{\theta}} \mathbb{E}[\psi(\bm{\theta}; d_i)]\Big|_{\bm{\theta} = \bm{\theta}_0}\,,$$
and
$$\bm{Q}_{\psi}(\bm{\theta}_0) =  \mathbb{E}[\psi(\bm{\theta}_0; d_i)\psi(\bm{\theta}_0; d_i)^T].$$\\
Assuming one can define an appropriate M-Estimator for a given problem, it would be possible to use these properties to perform statistical inference thereby allowing for different parametric tests.

With respect to the use of M-Estimation for the purposes of differential privacy, as mentioned, in this work we aim to integrate the robust $\rho(\cdot)$ functions within the OPM. More specifically, the OPM requires computing bounds on the first and second derivatives of the objective (loss) function $\rho(\cdot)$ so that the adequate amount of noise can be added to this objective function to ensure privacy. In order to compute these bounds we first propose a new specific function $\rho(\cdot)$ with bounded derivative $\psi(\cdot)$, delivering a robust M-Estimator that relies on a certain tuning constant. When this M-Estimator is used within the OPM, the tuning constant plays a role in determining the bounds of the above-mentioned derivatives and, consequently, plays a role in the amount of noise added for privacy. The following sections present the proposed bounded functi, i.e., RobHyt, and the statistical properties of the resulting non-private M-Estimator (Section \ref{sec_robhyt}) and, based on this, we then integrate this estimator within the OPM (Section \ref{sec_pert_mest}) to obtain the proposed Perturbed M-Estimation mechanism.

\subsection{The Robust Hyperbolic Tangent Function}
\label{sec_robhyt}

The $\rho$-functions that are usually employed for robustness purposes are either non-convex (e.g., Tukey Biweight) or are piecewise (and/or non continuously differentiable) functions (e.g., Huber) which make them either unusable within the OPM or can make the computation of the required sensitivity bounds and/or asymptotic properties more complicated. There exist other smooth (and strongly convex) functions, such as the Pseudo-Huber loss function but, given similar complexities in computing sensitivity bounds, we choose to address these issues by adapting the Hyperbolic Tangent ($\tanh$) function (see \cite{hampel1981change}) to deliver a bounded function for M-Estimation. The $\tanh$ function has nice properties since it is (i) continuously differentiable, (ii) defined over the entire real line and (iii) bounded between $[-1, 1]$ making it a good candidate for robustness purposes and for the derivation of the required sensitivity measures for the OPM. Given these properties, in Definition \ref{def_robhyt} we propose to modify this function by parametrizing it with a tuning constant $k \in \mathbb{R}^+$ that guarantees robustness when $k < \infty$ and converges towards the $L_2$-loss function when $k \to \infty$, similarly to the Huber loss-function. To the best of our knowledge, although various modifications of the hyperbolic functions have already been proposed and used for robust optimization (e.g., see  \cite{chen2017robust} and \cite{shen2019another}, to cite some recent work), we are not aware of a similar parametrization of this function in either the statistical or computer science literature to date.

\begin{definition}
\label{def_robhyt}
The Robust Hyperbolic Tangent (RobHyt) function is defined as follows
\begin{equation*}
    \rho_k(z) := \frac{k^2}{2} \log\left(\cosh\left(\frac{2}{k} z\right)\right),
\end{equation*}
where $k \in \mathbb{R}^+$.
\end{definition}
\noindent By definition, the proposed RobHyt function is convex with respect to its argument and has derivative given by
$$\psi_k(z) := k \tanh\left(\frac{2}{k} z\right),$$
which is bounded between $[-k, k]$. Hence, this function can be employed as a bounded function for robust M-Estimation since, as long as we choose $k < \infty$, we have that $\psi_k(z)$ is bounded and consequently so is the IF of the resulting M-estimator.

\begin{remark}
\label{k_lim}
The RobHyt function has the following important property: 
$$\underset{k \to \infty}{\lim} \rho_k(z) = z^2 .$$
\end{remark}

\noindent Given the above definition and remark,  this function can be seen as a smooth and differentiable-everywhere version of the Huber loss-function (similarly to the Pseudo-Huber loss). Keeping this in mind, we next consider an M-Estimator based on the commonly used $L_2$-loss function, i.e.,
\begin{equation}
\label{l2_loss}
	\tilde{\bm{\theta}} = \underset{\bm{\theta} \in \bm{\Theta}}{\argmin} \frac{1}{n}\sum_{i = 1}^n s(\bm{\theta}; d_i)^2,
\end{equation}
where $s(\bm{\theta}; d_i)$ is a score function such that under the true model we have that $\mathbb{E}[s(\bm{\theta}; d_i)]~=~0$.  An example is given by
\begin{equation*}
	s(\bm{\theta}; d_i) := y_i - \eta (x_i^T\bm{\theta}),
\end{equation*}
which represents the non-scaled score function for a Generalized Linear Model (GLM) where $y_i$ represents the response variable, $x_i \in \mathbb{R}^p$ a vector of covariates and $\eta(\cdot)$ a link function defined by the family characterizing the appropriate GLM model (see \cite{nelder1972generalized} and \cite{cantoni2001robust}). If we plug this score function, or any MLE score function corresponding to the derivative of the log-likelihood function, into (\ref{l2_loss}), then it is straightforward to see that the estimator $\tilde{\bm{\theta}}$ corresponds to the MLE. This definition is particularly relevant since, based on Remark \ref{k_lim}, it is also straightforward to see that the proposed M-Estimator
\begin{equation}
\label{tanh_est}
	\hat{\bm{\theta}} = \underset{\bm{\theta} \in \bm{\Theta}}{\argmin} \frac{1}{n}\sum_{i = 1}^n \rho_k(s(\bm{\theta}; d_i)),
\end{equation}
tends to the MLE as $k \to \infty$, in the same way as the Huber loss-function. 

In the robust statistical framework one chooses a fixed tuning constant $k$ based on the desired level of robustness and asymptotic efficiency with respect to the standard (non-robust) estimator. To do so, one usually requires an estimate of scale for the score function $s(\bm{\theta}; d_i)$ which could eventually be also obtained in a differentially private manner. If we let $k$ diverge with $n$, thereby defining the sequence $k_n \in \mathbb{R}^+$, the estimator in (\ref{tanh_est}) will inherit all the optimal properties of the MLE in terms of statistical accuracy based on the following assumptions (see \cite{newey1994large}):
\begin{enumerate}[label=\bfseries (A\arabic*), leftmargin = 1.5cm]
	\item The parameter space $\bm{\Theta}$ is compact. \label{cond_compact}
	\item $\mathbb{E}[s(\bm{\theta}; d_i)^2]$ is uniquely minimized in $\bm{\theta}_0$. \label{cond_ident}
	\item $\mathbb{E}[s(\bm{\theta}; d_i)^2]$ is continuous. \label{cond_cont}
	\item $\nicefrac{1}{n} \sum_{i=1}^n \rho_k(s(\bm{\theta}; d_i))$ converges uniformly in probability to $\mathbb{E}[s(\bm{\theta}; d_i)^2]$. \label{cond_uniconv}
\end{enumerate}
While assumption \ref{cond_compact} is a standard regularity condition which can eventually be replaced by other (model-specific) constraints, assumptions \ref{cond_ident} and \ref{cond_cont} are generally verified when considering the MLE. We now state our key result on the statistical consistency of the proposed estimator in (\ref{tanh_est}).

\begin{prop}
\label{prop.consistency}
Under assumptions \ref{cond_compact} to \ref{cond_cont} and assuming $s(\bm{\theta}; d_i) = \mathcal{O}_p(1)$, for all $k_n \in \mathbb{R}^+$ such that $k_n \to \infty$ as $n \to \infty$ we have that
$$\hat{\bm{\theta}} \overset{\mathcal{P}}{\to} \bm{\theta}_0 .$$
\end{prop}

\noindent This result, whose proof can be found in Appendix \ref{app.prop}, implies that as long as $k_n$ diverges at any given rate with $n$, the proposed estimator in (\ref{tanh_est}) is statistically consistent and hence converges in probability towards the true parameter $\bm{\theta}_0$. If however we assume that the score function $s(\bm{\theta}; d_i)$ is \textit{symmetrically} distributed, the following corollary delivers the convergence rate for a tuning constant $k_n \to 0$ with $n \to \infty$.

\begin{corol}
\label{corol_sym}
Let $x_n \in \mathbb{R}^+$ be a deterministic sequence such that $x_n \to 0$ and $\sqrt{n}\,x_n \to \infty$ as $n \to \infty$. Then, assuming $s(\bm{\theta}; d_i)$ has a symmetric distribution function and for any $k_n \geq x_n$, we have
$$\hat{\bm{\theta}} \overset{\mathcal{P}}{\to} \bm{\theta}_0 .$$
\end{corol}

\noindent The proof of this corollary is in Appendix \ref{app.cor}. These results are important since they allow us to define a region, which depends on the sample size $n$, within which we should define the tuning constant $k_n$ in order for our estimator $\hat{\bm{\theta}}$ to be statistically consistent. 

\begin{remark}
As stated earlier, for the purposes of robust statistical analysis the ``original'' tuning constant $k$ should be fixed and chosen, for example, with respect to the desired level of robustness and asymptotic efficiency of the resulting estimator compared to the non-robust alternative. However, for the purposes of privacy we would require the constant to be chosen also with respect to the sample size and noise for privacy, in addition to the asymptotic efficiency. Therefore, given the above results, we want to define a tuning constant $k_n$ that grows as slowly as possible since we want the statistical efficiency (low sampling variability) to dominate the noise added for differential privacy (which grows with $k_n$). A candidate could, for example, be $k_n := \log(\log(n))$ or any slowly increasing function in $n$. However, if we assume that the score function is (approximately) symmetrically distributed (e.g., linear regression with Gaussian residuals or logistic regression with probability $\pi \approx 0.5$) one could define, for example, $k_n := \nicefrac{1}{log(n)}$ for $n > 1$. At the same time however, a $k_n$ that is too small, despite allowing for consistency, can deliver an excessively inefficient estimator from a statistical point of view. Therefore a rule for determining $k_n$ based on the (asymptotic) efficiency under the constraint of consistency would be more appropriate and is left for future research.
\end{remark}

The next section explores the use of the above proposed and studied  M-Estimator within a differentially private mechanism in order to understand if the use of a robust M-Estimation framework can improve the utility of DP outputs 
for the same level of privacy. We also investigate the impact of the tuning constant $k_n$. For the purposes of notation, hereinafter we will simply denote the tuning constant as $k$ and make its underlying dependence on $n$ implicit whenever we let this constant diverge (or converge to zero).

\subsection{Tuned Objective Perturbation}
\label{sec_pert_mest}

In this section we propose the {\em Perturbed M-Estimation} mechanism which integrates the presented M-Estimator with the OPM framework; see Algorithm  \ref{KiferAlgorithm}. The reason for considering the OPM as a good candidate for integration with the above described M-Estimation framework is that the OPM, being the result of an empirical risk minimization problem, produces an output that can indeed be classified as an M-Estimator as in (\ref{def_mest}). Following the definition in \cite{kifer2012private}, the OPM for $\epsilon$-differential privacy is defined as follows
\begin{equation}
\label{eq_objpert}
	\bar{\bm{\theta}}_{DP} = \underset{\bm{\theta} \in \Theta}{\argmin} \frac{1}{n}\sum_{i = 1}^n l(\bm{\theta}; d_i) + \frac{\Delta}{2 n}\|\bm{\theta}\|_2^2 + \frac{b^T \bm{\theta}}{n},
\end{equation}
where $l(\bm{\theta}; d_i)$ is a convex loss function, $\Delta \geq \nicefrac{2 \lambda}{\epsilon}$, $\lambda$ is an upper bound on the eigenvalues of the Hessian $\nabla^2\, l(\bm{\theta}; d_i)$ and $b \in \mathbb{R}^p$ is a random vector with density
$$f(b) \propto \exp^{-\epsilon \nicefrac{\|b\|_2}{2 \xi}},$$
where $\xi$ is such that $\|\nabla\, l(\bm{\theta}; d_i)\|_2 \leq \xi$. Therefore $\lambda$ and $\xi$ are two parameters that define the sensitivity measures of the loss function and consequently impact the amount of noise (perturbation) that is added to the loss function. Considering the defintion in (\ref{eq_objpert}), we can now replace the loss function $l(\bm{\theta}; d_i)$ with the proposed loss function in (\ref{tanh_est}) to deliver the new Perturbed M-Estimator.

\begin{definition}
\label{def_pert_mest}
The Perturbed M-Estimator is defined as follows
\begin{equation}
\label{eq_pertmest}
	\hat{\bm{\theta}}_{DP} = \underset{\bm{\theta} \in \Theta}{\argmin} \frac{1}{n}\sum_{i = 1}^n \rho_k(s(\bm{\theta}; d_i)) + \frac{\Delta_k}{2 n}\|\bm{\theta}\|_2^2 + \frac{b_k^T \bm{\theta}}{n}.
\end{equation}
\end{definition}

\noindent From the above definition, we have that $\Delta_k$ and $b_k$ (which depend on $\lambda_k$ and $\xi_k$ respectively) are now quantities and variables that depend on the tuning constant $k$. Indeed, we have that $\xi_k$, and hence $b_k$, depends on the following quantity
\begin{equation*}
	\nabla \rho_k(s(\bm{\theta}; d_i)) = \underbrace{\tanh\left(\frac{2}{k} s(\bm{\theta}; d_i)\right)}_{\in [-1, 1]} \, k \, \nabla s(\bm{\theta}; d_i),
\end{equation*}
while $\lambda_k$, and hence $\Delta_k$, depends on
\begin{equation*}
	\nabla^2 \rho_k(s(\bm{\theta}; d_i)) = 2 \underbrace{\text{sech}\left(\frac{2}{k} s(\bm{\theta}; d_i)\right)^2}_{\in [0, 1]} \left(\nabla s(\bm{\theta}; d_i)\right)^2 + \underbrace{\tanh\left(\frac{2}{k} s(\bm{\theta}; d_i)\right)}_{\in [-1, 1]} \, k \, \nabla^2 s(\bm{\theta}; d_i) .
\end{equation*}
From the above expressions we observe that the tuning constant $k$ can be directly related to a specific notion of DP-based sensitivity for $\nabla \, s(\bm{\theta}; d_i)$ and $\nabla^2 \, s(\bm{\theta}; d_i)$. Based on these expressions, for example, one could choose to define the tuning constant $k$ as being inversely proportional to the sensitivity of these expressions according to the problem at hand.

Our proposed approach, highlighted in Algorithm \ref{KiferAlgorithm}, can therefore be seen as a form of ``tuned'' objective perturbation where we can calibrate the choice of $k$ based on (i) sample size, (ii) required statistical efficiency and (iii) known sensitivity bounds for the loss function. Indeed, we would generally want to choose a $k$ that is ``small'' to achieve low sensitivity bounds (and add less noise for privacy) but, in order to achieve statistical efficiency, we would ideally want $k$ not to be \textit{too} small. As stated earlier, the study of an optimal (private) choice of the tuning constant is left for future research.

\vspace{0.4cm}

\begin{algorithm}
\caption{Perturbed M-Estimation -- modified Objective Perturbation from \cite{kifer2012private}}
\scriptsize
  \vspace{0.1cm}
INPUT: $D\in \mathcal{D}^n$, $\epsilon>0$, a tuning parameter $k \in \mathbb{R}^+$, a convex set $\Theta \subset \mathbb{R}^p$, a convex loss  $\hat{L}_k(\bm{\theta}; D) =\frac1n \sum_{i=1}^n \ \rho_k(s(\bm{\theta}; d_i))$ defined on $\Theta$ such that the Hessian $\nabla^2 \rho_k(s(\bm{\theta}; d))$ is continuous in  $\bm{\theta}$ and $d$, $\xi_k>0$ such that $\lVert \nabla \rho_k(s(\bm{\theta}; d))\rVert_2\leq \xi_k$ for all $\bm{\theta} \in \Theta$ and $d\in D$, and $\lambda_k>0$  is an upper bound on the eigenvalues of $\nabla^2\rho_k(s(\bm{\theta}; d))$ for all $\bm{\theta} \in \Theta$ and $d\in D$.\\
\begin{algorithmic}[1]
  \setlength\itemsep{0em}
  \STATE Set $\Delta_k = \frac{2\lambda_k}{\epsilon}$
  \vspace{0.1cm}
\STATE Draw $b_k\in \mathbb{R}^m$ from the density $f(b_k;\epsilon, \xi)\propto \exp(-\frac\epsilon{2\xi}\lVert b_k\rVert_2)$
  \vspace{0.1cm}
\STATE Compute $\hat{\bm{\theta}}_{DP} = \underset{\bm{\theta} \in \Theta}{\argmin} \frac{1}{n}\sum_{i = 1}^n \rho_k(s(\bm{\theta}; d_i)) + \frac{\Delta_k}{2 n}\|\bm{\theta}\|_2^2 + \frac{b_k^T \bm{\theta}}{n}$\\
\end{algorithmic}
  \vspace{0.1cm}
OUTPUT: $\hat{\bm{\theta}}_{DP}$
\label{KiferAlgorithm}
\end{algorithm}

\section{Applications and Simulations}
\label{sec.app_sim}

In this section we investigate the potential utility of the suggested approach in Algorithm \ref{KiferAlgorithm} in some applied and simulated settings. The examples are based on standard linear regression and logistic regression for small and large sample sizes and with a guarantee for pure differential privacy with $\epsilon = 0.1$. The parameter of interest $\bm{\theta}$ is represented by the regression coefficient vector $\bm{\beta} \in \mathbb{R}^p$ and the utility of the estimators is measured via the $L_2$-norm (i) between the estimators and the reference value (non-private estimator or true value) or (ii) between the observed response and the predictions based on the different estimators (mean squared prediction error). The performance of the estimators is evaluated over $H = 100$ replications and, for each of them, different values of the tuning constant $k$ are considered between $[0, 2]$. The latter range is considered since if $k < 1$, then the sensitivity measures for privacy are reduced while for values $k > 1$ the sensitivity is increased.

\begin{remark}
It must be noted that the OPM (and hence the proposed estimator) requires the optimization procedure to converge in order to guarantee differential privacy. In few examples we did not have the convergence, but we still included them in the overall results to illustrate the potential gains in utility that this new approach could deliver. Hence, the results in this section should be considered as preliminary investigations rather than ``conclusive'' empirical results. Based on these observations, the goal would be to explore possibly more numerically stable privacy mechanisms for the considered approach using, for example, the stochastic gradient descent method (see e.g., \cite{rajkumar2012differentially}, \cite{song2013stochastic}, \cite{wang2015privacy}, \cite{chen2019renyi}) or the more recent KNG approach proposed by \cite{reimherr2019kng}.
\end{remark} 

\subsection{Applications: Linear Regression}

For the linear regression examples, let $\bm{y} \in \mathbb{R}^n$ be a vector of responses and $\bm{X} \in \mathbb{R}^{n \times p}$ be a matrix of covariates, where the first column is a vector of ones for the intercept term. The score function is given by
$$s(\bm{\beta}; d_i) := y_i - x_i^T\bm{\beta} ,$$
where $y_i$ is the response variable and $x_i \in \mathbb{R}^p$ is the vector of covariates for the $i^{th}$ row. We compare the following estimators:
\begin{itemize}
	\item $\hat{\bm{\beta}} = (\bm{X}^T\bm{X})^{-1}\bm{X}^T\bm{y}$: the least-squares non-private estimator that will be used as reference for the other estimators (i.e., considered as the true $\beta$ we aim to estimate).
    \item $\bar{\bm{\beta}}$: the non-private robust estimator using $\rho_k(s(\bm{\beta}; d_i))$; we expect this to converge to $\hat{\bm{\beta}}$ as $k \to \infty$.
    \item $K$-norm Sufficient Statistics Perturbation: this approach is proposed by \cite{awan2018structure} and delivers differentially private estimators based on different norms considered for the sensitivity of the sufficient statistics $\bm{X}^T\bm{X}$ and $\bm{X}^T\bm{y}$ (see the functional mechanism of  \cite{zhang2012functional}). Based on these norms (including an optimal $K$-norm defined in \cite{awan2018structure}), appropriate noise is added to the sufficient statistics to deliver differentially private estimators based on replacing an observation as opposed to removing an observation.
    \item $\tilde{\bm{\beta}}$: the proposed Perturbed M-Estimator in (\ref{eq_pertmest}). 
\end{itemize}

\begin{figure}

\begin{center}
\begin{tikzpicture}[scale=0.9]

    \begin{axis}[
    xlabel={Tuning Constant $k$},
    ylabel={$\log(\|\hat{y}^* - y\|_2)$},
    xmin=0, xmax=2,
    ymin=-3.1, ymax=9.9,
    xtick={0, 0.5, 1, 1.5, 2},
    ytick={-2, 0, 2, 4, 6, 8},
    legend style={at={(0.9,0.9)}, anchor=north east}, 
    ymajorgrids=true,
    grid style=dashed,
]
 
\addplot[mark=none, color=blue, ultra thick]
    coordinates {
    (0.01,3.6972) (0.1147,3.6972) (0.2195,3.6972) (0.3242,3.6972) (0.4289,3.6972) (0.5337,3.6972) (0.6384,3.6972) (0.7432,3.6972) (0.8479,3.6972) (0.9526,3.6972) (1.0574,3.6972) (1.1621,3.6972) (1.2668,3.6972) (1.3716,3.6972) (1.4763,3.6972) (1.5811,3.6972) (1.6858,3.6972) (1.7905,3.6972) (1.8953,3.6972) (2,3.6972) 
    };
    \addlegendentry{$K$-Norm}
    
    \addplot[mark=none, color=orange, ultra thick]
    coordinates {
    (0.01,9.1768) (0.1147,9.1768) (0.2195,9.1768) (0.3242,9.1768) (0.4289,9.1768) (0.5337,9.1768) (0.6384,9.1768) (0.7432,9.1768) (0.8479,9.1768) (0.9526,9.1768) (1.0574,9.1768) (1.1621,9.1768) (1.2668,9.1768) (1.3716,9.1768) (1.4763,9.1768) (1.5811,9.1768) (1.6858,9.1768) (1.7905,9.1768) (1.8953,9.1768) (2,9.1768) 
    };
    \addlegendentry{$L_1$-Norm}
    
\addplot[mark=none, color=purple, ultra thick]
    coordinates {
    (0.01,7.9953) (0.1147,7.9953) (0.2195,7.9953) (0.3242,7.9953) (0.4289,7.9953) (0.5337,7.9953) (0.6384,7.9953) (0.7432,7.9953) (0.8479,7.9953) (0.9526,7.9953) (1.0574,7.9953) (1.1621,7.9953) (1.2668,7.9953) (1.3716,7.9953) (1.4763,7.9953) (1.5811,7.9953) (1.6858,7.9953) (1.7905,7.9953) (1.8953,7.9953) (2,7.9953) 
    };
    \addlegendentry{$L_{\infty}$-Norm}

    \addplot[mark=none, color=red, ultra thick]
    coordinates {
    (0.01,-2.4321) (0.1147,-2.547) (0.2195,-2.5642) (0.3242,-2.5689) (0.4289,-2.5638) (0.5337,-2.5767) (0.6384,-2.5782) (0.7432,-2.5792) (0.8479,-2.5794) (0.9526,-2.5796) (1.0574,-2.5806) (1.1621,-2.571) (1.2668,-2.5802) (1.3716,-2.5808) (1.4763,-2.5806) (1.5811,-2.5793) (1.6858,-2.5808) (1.7905,-2.5803) (1.8953,-2.5809) (2,-2.5806) 
    };
    \addlegendentry{$\bar{\bm{\beta}}$}
    
    \addplot[mark=none, color=green, ultra thick]
    coordinates {
    (0.01,-1.2308) (0.1147,-1.2325) (0.2195,-1.2227) (0.3242,-1.1993) (0.4289,-1.1706) (0.5337,-1.1167) (0.6384,-1.0702) (0.7432,-1.0175) (0.8479,-0.9587) (0.9526,-0.9021) (1.0574,-0.8338) (1.1621,-0.759) (1.2668,-0.694) (1.3716,-0.6219) (1.4763,-0.5552) (1.5811,-0.4737) (1.6858,-0.4031) (1.7905,-0.3537) (1.8953,-0.2663) (2,-0.1992) 
    };
    \addlegendentry{$\tilde{\bm{\beta}}$}

\end{axis}

\end{tikzpicture}
\begin{tikzpicture}[scale=0.9]

    \begin{axis}[
    xlabel={Tuning Constant $k$},
    ylabel={$\log(\|\beta^* - \hat{\beta}\|_2)$},
    xmin=0, xmax=2,
    ymin=-1.3, ymax=2.5,
    xtick={0, 0.5, 1, 1.5, 2},
    ytick={-1, 0, 1, 2},
    legend style={at={(0.9,0.9)}, anchor=north east}, 
    ymajorgrids=true,
    grid style=dashed,
]
 
\addplot[mark=none, color=blue, ultra thick]
    coordinates {
    (0.01,1.8612) (0.2311,1.8612) (0.4522,1.8612) (0.6733,1.8612) (0.8944,1.8612) (1.1156,1.8612) (1.3367,1.8612) (1.5578,1.8612) (1.7789,1.8612) (2,1.8612) 
    };
    \addlegendentry{$K$-Norm}

\addplot[mark=none, color=orange, ultra thick]
    coordinates {
    (0.01,2.2941) (0.2311,2.2941) (0.4522,2.2941) (0.6733,2.2941) (0.8944,2.2941) (1.1156,2.2941) (1.3367,2.2941) (1.5578,2.2941) (1.7789,2.2941) (2,2.2941) 
    };
    \addlegendentry{$L_1$-Norm}
    
    \addplot[mark=none, color=purple, ultra thick]
    coordinates {
    (0.01,1.7931) (0.2311,1.7931) (0.4522,1.7931) (0.6733,1.7931) (0.8944,1.7931) (1.1156,1.7931) (1.3367,1.7931) (1.5578,1.7931) (1.7789,1.7931) (2,1.7931)
    };
    \addlegendentry{$L_{\infty}$-Norm}

    \addplot[mark=none, color=red, ultra thick]
    coordinates {
    (0.01,-0.6299) (0.2311,-0.5644) (0.4522,-0.6529) (0.6733,-0.8187) (0.8944,-0.3249) (1.1156,-1.0272) (1.3367,-0.6081) (1.5578,-1.0745) (1.7789,-0.6991) (2,-0.7117) 
    };
    \addlegendentry{$\bar{\bm{\beta}}$}
    
    \addplot[mark=none, color=green, ultra thick]
    coordinates {
    (0.01,-0.218) (0.2311,-0.3941) (0.4522,-0.4336) (0.6733,-0.4628) (0.8944,-0.4615) (1.1156,-0.424) (1.3367,-0.3885) (1.5578,-0.358) (1.7789,-0.3028) (2,-0.2561) 
    };
    \addlegendentry{$\tilde{\bm{\beta}}$}

\end{axis}

\end{tikzpicture}
\end{center}

\caption{Left: Logarithm of the mean squared prediction error versus tuning constant $k$ for the attitude dataset. Right: Logarithm of the mean squared error versus tuning constant $k$ with respect to the non-private estimator $\hat{\bm{\beta}}$ for the housing dataset.} \label{fig:linear}

\end{figure}
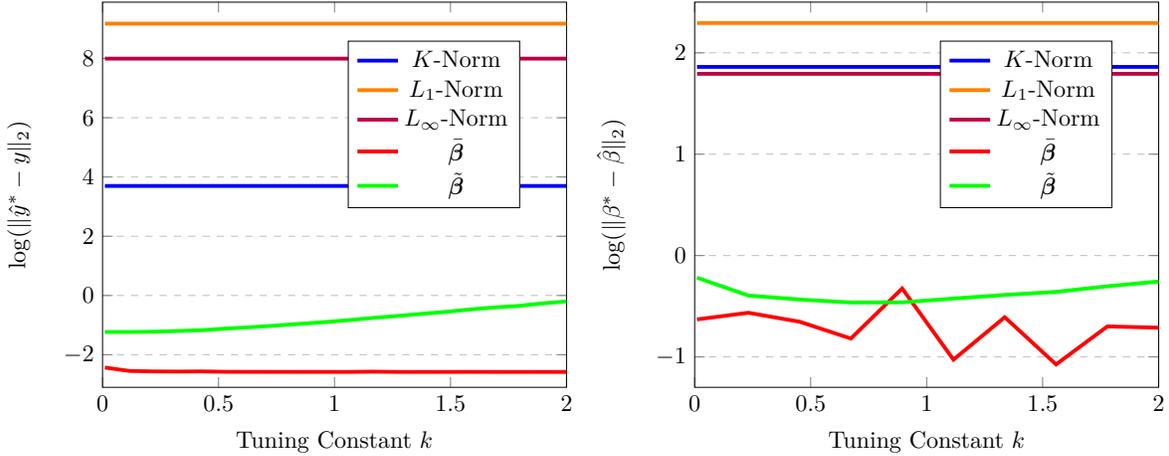

The first example we use is the ``Attitude'' dataset from \cite{chatterjee2015regression}, available in the \texttt{R} statistical software. This is a small dataset with only 30 observations and 7 variables capturing the percentages of favorable responses to a survey of clerical employees in a financial organization. A question of interest is how each variable contributes to the overall rating of the company ($\bm{y}$). The left plot in Figure \ref{fig:linear} reports the mean square prediction error for the different estimators over the different values of the tuning constant $k$ for this data. 

The second example, the ``San Francisco housing" data, has been used for the evaluation of different statistical and differentially private methods; we use a dataset version from  \cite{awan2018structure}. This dataset consists of  $348,189$ observations on houses in the Bay area between 2003 and 2006. The main question of  interest is in explaining the rent of the houses as a function of several other variables (e.g., square-footage, location, age of house, number of bedrooms, county). The right plot in Figure \ref{fig:linear} shows the mean squared error between all estimators and the parameter of reference $\hat{\bm{\beta}}$ (the non-private estimator). 

In both cases, the data are pre-processed by taking the logarithm of some numerical variables and ensuring that all numerical variables lie between $[-1, 1]$. The latter bounding is not necessarily required for Perturbed M-Estimation since the tuning constant $k$ can eventually compensate for a higher sensitivity due to larger bounds on the variables; nevertheless we perform this processing in order to compare it with the other estimators.

From Figure~\ref{fig:linear}, we can notice that the only estimators that depend on the tuning constant $k$ (and whose lines therefore do not remain constant) are the robust non-private estimator $\bar{\bm{\beta}}$ and the proposed DP  $\tilde{\bm{\beta}}$. For both datasets, as expected, it is clear that the robust non-private estimator $\bar{\bm{\beta}}$ (red line) has the best performance, and as the tuning constant $k$ increases, this estimator improves its performance since it will converge to $\hat{\bm{\beta}}$. Our proposed DP M-estimator (green line) appears to be the best alternative, and it significantly outperforms the other DP estimators in these settings. However, for both datasets, the performance of the Perturbed M-estimator gets worse as the value of the tuning constant increases (although it still does better than the other DP estimators).  This implies that the noise added for privacy starts to dominate over the statistical efficiency that is delivered through the increase of the tuning constant. Another effect that is more evident for the housing data (right plot) is that the performance of $\tilde{\bm{\beta}}$ is not optimal for the smallest values of $k$ since it decreases and then starts to steadily increase around $k = 1$. This would indicate that for small values of $k$, the statistical inefficiency dominates the minimal noise added for privacy, while as $k$ increases, this ratio starts to diminish as a  result of the increasing statistical efficiency being overcome by the noise added for privacy.

\subsection{Simulations: Logistic Regression}

The simulation study in this section replicates the one in \cite{awan2018structure} but with a smaller sample size of $n = 100$. We consider a logistic regression model where we generate uniformly distributed covariates $x_i \sim \mathcal{U}[-1, 1]$ and set the true parameter vector as
$\bm{\beta} = (0, -1, \nicefrac{-1}{2}, \nicefrac{-1}{4}, 0, \nicefrac{3}{4}, \nicefrac{3}{2}).$
Based on these values, we simulate uniform values $U_i \sim \mathcal{U}[0, 1]$ and define the simulated response values using the link function
$$\eta (x_i^T\bm{\beta}) = \frac{\exp{(x_i^T\bm{\beta})}}{1 + \exp{(x_i^T\bm{\beta})}} ,$$
as follows
\[ y_i = \begin{cases} 
      1 & U_i < \eta(x_i^T\bm{\beta}) \\
      0 & \text{otherwise}.
   \end{cases}
\]
In this case, the score function is defined as $s(\bm{\beta}; d_i) = y_i - \eta(x_i^T\bm{\beta})$. In order for the robust non-private estimator to be Fisher consistent we would need to derive a correction term since the bounded function can introduce bias in the resulting estimator (see e.g., \cite{cantoni2001robust}). For the purpose of this simulation we do not apply this correction since given our setting, the scores are approximately symmetrically distributed. Moreover, we assume that the performance of the proposed approach can only be improved if the correction term was introduced (and would be less relevant when $k \to \infty$).

\begin{figure}
\begin{center}
\begin{tikzpicture}

    \begin{axis}[
    xlabel={Tuning Constant $k$},
    ylabel={$\log(\|\beta^* - \beta\|_2)$},
    xmin=0, xmax=2,
    ymin=-0.1, ymax=3.8,
    xtick={0, 0.5, 1, 1.5, 2},
    ytick={0, 1, 2, 3},
    legend style={at={(0.05,0.95)}, anchor= north west}, 
    ymajorgrids=true,
    grid style=dashed,
]

\addplot[mark=none, color=red, ultra thick]
    coordinates {
    (0.01,0.1442) (0.1147,0.1442) (0.2195,0.1442) (0.3242,0.1442) (0.4289,0.1442) (0.5337,0.1442) (0.6384,0.1442) (0.7432,0.1442) (0.8479,0.1442) (0.9526,0.1442) (1.0574,0.1442) (1.1621,0.1442) (1.2668,0.1442) (1.3716,0.1442) (1.4763,0.1442) (1.5811,0.1442) (1.6858,0.1442) (1.7905,0.1442) (1.8953,0.1442) (2,0.1442) 
    };
    \addlegendentry{$\hat{\bm{\beta}}$}

\addplot[mark=none, color=blue, ultra thick]
    coordinates {
    (0.01,3.3025) (0.1147,3.3025) (0.2195,3.3025) (0.3242,3.3025) (0.4289,3.3025) (0.5337,3.3025) (0.6384,3.3025) (0.7432,3.3025) (0.8479,3.3025) (0.9526,3.3025) (1.0574,3.3025) (1.1621,3.3025) (1.2668,3.3025) (1.3716,3.3025) (1.4763,3.3025) (1.5811,3.3025) (1.6858,3.3025) (1.7905,3.3025) (1.8953,3.3025) (2,3.3025) 
    };
    \addlegendentry{$L_1$-Norm}

\addplot[mark=none, color=purple, ultra thick]
    coordinates {
    (0.01,3.0772) (0.1147,3.0772) (0.2195,3.0772) (0.3242,3.0772) (0.4289,3.0772) (0.5337,3.0772) (0.6384,3.0772) (0.7432,3.0772) (0.8479,3.0772) (0.9526,3.0772) (1.0574,3.0772) (1.1621,3.0772) (1.2668,3.0772) (1.3716,3.0772) (1.4763,3.0772) (1.5811,3.0772) (1.6858,3.0772) (1.7905,3.0772) (1.8953,3.0772) (2,3.0772) 
    };
    \addlegendentry{$L_{2}$-Norm}
    
    \addplot[mark=none, color=black, ultra thick]
    coordinates {
    (0.01,2.5855) (0.1147,2.5855) (0.2195,2.5855) (0.3242,2.5855) (0.4289,2.5855) (0.5337,2.5855) (0.6384,2.5855) (0.7432,2.5855) (0.8479,2.5855) (0.9526,2.5855) (1.0574,2.5855) (1.1621,2.5855) (1.2668,2.5855) (1.3716,2.5855) (1.4763,2.5855) (1.5811,2.5855) (1.6858,2.5855) (1.7905,2.5855) (1.8953,2.5855) (2,2.5855) 
    };
    \addlegendentry{$L_{\infty}$-Norm}
    
    \addplot[mark=none, color=orange, ultra thick]
    coordinates {
    (0.01,1.0871) (0.1147,1.0871) (0.2195,1.0871) (0.3242,1.0871) (0.4289,1.0871) (0.5337,1.0871) (0.6384,1.0871) (0.7432,1.0871) (0.8479,1.0871) (0.9526,1.0871) (1.0574,1.0871) (1.1621,1.0871) (1.2668,1.0871) (1.3716,1.0871) (1.4763,1.0871) (1.5811,1.0871) (1.6858,1.0871) (1.7905,1.0871) (1.8953,1.0871) (2,1.0871) 
    };
    \addlegendentry{$L_{\infty}^*$-Norm}
    
    \addplot[mark=none, color=green, ultra thick]
    coordinates {
    (0.01,0.7087) (0.1147,0.712) (0.2195,0.7207) (0.3242,0.7325) (0.4289,0.7477) (0.5337,0.7663) (0.6384,0.7865) (0.7432,0.8077) (0.8479,0.8329) (0.9526,0.8596) (1.0574,0.8871) (1.1621,0.9146) (1.2668,0.9372) (1.3716,0.97) (1.4763,0.9978) (1.5811,1.0251) (1.6858,1.0526) (1.7905,1.081) (1.8953,1.1036) (2,1.1309) 
    };
    \addlegendentry{$\tilde{\bm{\beta}}$}

\end{axis}

\end{tikzpicture}
\end{center}

\caption{Logarithm of the mean squared error versus tuning constant $k$ with respect to the true parameter value $\bm{\beta}$ for the logistic regression simulation.} \label{fig:logistic}

\end{figure}
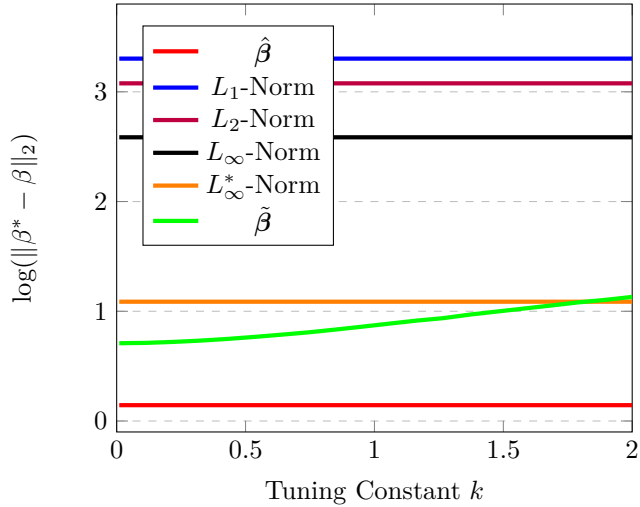

We consider the following estimators for this simulation study:
\begin{itemize}
    \item $\hat{\beta}$: MLE for logistic regression as a non-private reference.
    \item Objective perturbation estimators based on $K$-norms: private estimators based on the generalized OPM (adapted from \cite{awan2018structure}, and \cite{kifer2012private}) using different $K$-norms with change-DP, i.e., replace an individual. We consider the following norms, $L_1$, $L_2$, $L_{\infty}$ and another version of the $L_{\infty}$-based OPM with an additional tuning constant to control the bias-variance trade-off set to $q = 0.85$ (instead of $q = 0.5$ for the other estimators, see \cite{awan2018structure}).
    \item $\tilde{\bm{\beta}}$: the proposed Perturbed M-Estimator in (\ref{eq_pertmest}). 
\end{itemize}
The additional tuning constant $q$ and the use of other norms could also be considered for our proposed approach in order to improve its performance. However, for this paper, we keep it only depending on the tuning constant $k$. The mean squared errors with respect to the true parameter vector $\bm{\beta}$ are presented in Figure \ref{fig:logistic}.
 
The conclusions are similar to those of the previous section for the linear regression setting. Obviously, the MLE (red line) performs the best. However our proposed DP M-estimator (green line) is the best alternative, in some cases having substantially better performance than other DP estimators that are more commonly used with logistic regression.  We also see that in this case the performance of $\tilde{\bm{\beta}}$ appears to worsen more rapidly as $k$ increases but not much more than in the Attitude dataset; recall, both of these datasets are on a smaller scale with $n=30$ and $n=100$ --- settings where differenitally private mechanisms, in general have a harder time producing accurate statistics with small privacy-loss budgets, $\epsilon$. In addition, we can see that the private estimator based on the tuned $L_{\infty}$-norm ($L_{\infty}^*$) also has a high utility, as argued in \cite{awan2018structure}, and is very close to our proposed approach. Nevertheless, as alluded earlier, it is possible that our approach could also benefit from consideration of other norms and the additional tuning constant $q$; we leave that to future work.  

\section{Conclusions and Outlook}
\label{sec.conclusion}

In this work we consider the use of methods from the field of robust statistics in order to improve the utility of differentially private mechanisms, that is of their statistical outputs. More specifically, we propose a robust M-Estimator with well defined properties, including consistency, and propose to employ it within the popular objective perturbation mechanism, thereby proposing a Perturbed M-Estimation mechanism. Our approach allows for calibration of noise needed to produce differentially private estimates and it improves statistical utility of these outputs while removing the need to impose bounds on the parameter space and the response variables --- this is a significant methodological and practical contribution as many current mechanisms require pre-processing of data such that it is bounded.  There is still the need, however, to impose bounds on the covariates, for regression problems for example, in order to determine sensitivity bounds. Our preliminary simulations and examples for linear and logistic regressions demonstrate significantly improved utility in estimation of parameter estimates under $\epsilon$-DP in comparison to the currently used methods.  It is also important to note that our proposed DP estimator works reasonably well for small sample sizes $n$. The setting with small $n$ is frequently problematic for DP since the noise needed to protect the privacy may overcome the sampling noise too much, making data unusable. While the choice of the tuning constant $k$ for our robust estimator is more obvious in the non-private setting, and is tied to $n$, in the private setting the clear rules are yet to be determined, and are part of future work. 

Having investigated the possible use of robust statistical tools in the domain of differential privacy, it appears that it is worth to further explore this path and better understand properties and convergence rates of the proposed approach. Two improvements that can be considered jointly are the use of a Mallow's type estimator (see \cite{huber2011robust}, \cite{maronna2019robust}) and the redefinition of the expression for the OPM based on the properties of the proposed RobHyt function or of any other function with bounded derivative (and definable sensitivity bounds) and with similar properties of consistency. The Mallow's type estimator can automatically bound the covariates of a regression problem thereby possibly removing the need to impose any bounds on parameters and data. The redefining of the sensitivity bounds can be done, for example, by using the links between smooth sensitivity and the GES as highlighted in \cite{chaudhuri2012convergence} and \cite{avella2019privacy}). Moreover, depending on the definition of the problem, rules to determine the value of the tuning constant $k$ can be developed or appropriate methods to select an ``optimal'' $k$ in a private manner can be studied. In the latter case, an intuitive approach would be to find the value of $k$ based on the definition of the asymptotic variance for M-Estimators which would possibly depend only on the model and the sample size thereby allowing to determine it independently from the data (or find an approximation in a private manner). Another approach that will be worth investigating is the use of a private stochastic gradient descent mechanism (see \cite{song2013stochastic}, \cite{wang2015privacy}, \cite{chen2019renyi}), or methods such as the KNG mechanism in \cite{reimherr2019kng}, in order to overcome possible non-convergence issues of the objective perturbation mechanism. Finally, once possible new sensitivity bounds are defined based on robust statistical measures, it would be possible to deliver the corresponding statistical inference framework that would allow to construct private confidence intervals and perform private parametric tests. And, nearly fifty years ago after \cite{andrews1972etal} provided an extensive survey of some 68 robust estimates of location, we can take a look back at those in order to move forward. 

\vspace{1.5cm}

\noindent \textbf{Acknowledgements} The authors would like to thank Jordan Awan and Dan Kifer for the useful discussions and inputs as well as for sharing code to compare the results in this paper with existing approaches. We would like to thank St\'{e}phane Guerrier and Mucyo Karemera for their helpful suggestions, and Marco Avella-Medina for sharing his working manuscript with us. This research was supported in part by NSF Grants  SES-1534433 and SES-1853209 to Pennsylvania State University, and by the National Center for Advancing Translational Sciences, National Institutes of Health, through Grant UL1 TR002014. The content is solely the responsibility of the authors and does not necessarily represent the official views of the NIH or NSF. Part of this work was done while authors were visiting the Simons Institute for the Theory of Computing.

\newpage

\bibliographystyle{plainnat}
\bibliography{references}

\newpage
\appendix

\section{Proofs}

\subsection{Proof of Proposition \ref{prop.consistency}}
\label{app.prop}

\begin{proof}
Provided the other assumptions hold (which is generally common or verified when considering the MLE), we need to prove that Assumption \ref{cond_uniconv} holds as well. The definition of uniform convergence for our setting is the following
\begin{equation*}
	\underset{\bm{\theta} \in \bm{\Theta}}{\sup} \,\, \Big|\nicefrac{1}{n} \sum_{i=1}^n \rho_k(s(\bm{\theta}, d_i)) - \mathbb{E}[s(\bm{\theta}, d_i)^2]\Big| \to 0.
\end{equation*} 
Let us denote $Q_n(s(\bm{\theta})) := \nicefrac{1}{n} \sum_{i=1}^n \rho_k(s(\bm{\theta}, d_i))$ and $Q_0(s(\bm{\theta})) := \mathbb{E}[s(\bm{\theta}, d_i)^2]$. We can therefore re-express the above definition as
\begin{equation*}
	\underset{\bm{\theta} \in \bm{\Theta}}{\sup} \,\, \Big|Q_n(s(\bm{\theta})) - Q_0(s(\bm{\theta}))\Big| \to 0,
\end{equation*} 
where, defining $\bar{Q}_n(s(\bm{\theta})):= \nicefrac{1}{n} \sum_{i=1}^n s(\bm{\theta}, d_i)^2$, by triangle inequality we have
\begin{equation}
\label{eq_trianineq}
	\Big|Q_n(s(\bm{\theta})) - Q_0(s(\bm{\theta}))\Big| \leq \Big|Q_n(s(\bm{\theta})) -  \bar{Q}_n(s(\bm{\theta}))\Big| + \Big|\bar{Q}_n(s(\bm{\theta})) - Q_0(s(\bm{\theta}))\Big|.
\end{equation} 
Since we assume that $s(\bm{\theta}, d_i) = \mathcal{O}_p(1)$, we have that
$$\Big| \bar{Q}_n(s(\bm{\theta})) - Q_0(s(\bm{\theta}))\Big| = \mathcal{O}_p\left(\frac{1}{\sqrt{n}}\right) ,$$
based on the weak law of large numbers and Markov's inequality. We therefore focus on the first term on the right side of the inequality in (\ref{eq_trianineq}). For this reason, let us apply a second order Taylor expansion of the two functions characterizing this term around the expected value of $s(\bm{\theta}, d_i)$ at the solution (i.e. zero):
$$Q_n(s(\bm{\theta})) = \frac{1}{n}\sum_{i=1}^n \left[ \rho_k(0) + \frac{\partial}{\partial s_i(\bm{\theta})} \rho_k(s_i(\bm{\theta}))\Big|_{s_i(\bm{\theta}) = 0} s_i(\bm{\theta}) + \frac{\partial^2}{\partial^2 s_i(\bm{\theta})} \rho_k(s_i(\bm{\theta}))\Big|_{s_i(\bm{\theta}) = 0} s_i(\bm{\theta})^2 + R_{\rho_k}\right] ,$$
and
$$\bar{Q}_n(s(\bm{\theta})) = \frac{1}{n}\sum_{i=1}^n \left[ 0 + 2 \cdot 0 \cdot s_i(\bm{\theta}) + 2 s_i(\bm{\theta})^2 \right] = \frac{1}{n}\sum_{i=1}^n 2s_i(\bm{\theta})^2 ,$$
since there is no remainder term for the expansion of $\bar{Q}_n(s(\bm{\theta}))$. As for the expansion of $Q_n(s(\bm{\theta}))$, by taking the required derivatives and evaluating them in zero we end up with
$$Q_n(s(\bm{\theta})) = \frac{1}{n}\sum_{i=1}^n \left[ 2 \underbrace{\text{sech}(0)^2}_{=1} s_i(\bm{\theta})^2 + R_{\rho_k}\right] = \frac{1}{n}\sum_{i=1}^n \left[ 2 s_i(\bm{\theta})^2 + R_{\rho_k}\right] .$$
Considering these expansions, we have that
$$Q_n(s(\bm{\theta})) -  \bar{Q}_n(s(\bm{\theta})) = \frac{1}{n}\sum_{i=1}^n R_{\rho_k},$$
and therefore let us take a look at the remainder term that has the following structure
$$R_{\rho_k} = \frac{\partial^3}{\partial^3 s(\bm{\theta})} \rho_k(s(\bm{\theta})) \frac{s(\bm{\theta})^3}{3!}.$$
Let us focus on the bound of the third derivative and, taking the absolute value, we have
\begin{eqnarray*}
\Big| \frac{\partial^3}{\partial^3 s(\bm{\theta})} \rho_k(s(\bm{\theta})) \Big| &= \Big| \frac{6}{k}\tanh\left(\frac{2}{k}s(\bm{\theta})\right)\text{sech}\left(\frac{2}{k}s(\bm{\theta})\right) \Big| \\
&= \frac{6}{k} \underbrace{\Big| \tanh\left(\frac{2}{k}s(\bm{\theta})\right) \Big|}_{\leq 1} \underbrace{\Big| \text{sech}\left(\frac{2}{k}s(\bm{\theta})\right) \Big|}_{\leq 1} \leq \frac{6}{k} .
\end{eqnarray*}
Hence, we have that
$$\Big|R_{\rho_k}\Big| \leq \frac{6}{k} \frac{s(\bm{\theta})^3}{3!} = \frac{s(\bm{\theta})^3}{k} ,$$
which, for $k \to \infty$ with $n$, implies that
$$\Big| Q_n(s(\bm{\theta})) -  \bar{Q}_n(s(\bm{\theta})) \Big| = \mathcal{O}_p\left(\frac{1}{k}\right),$$
since $s(\bm{\theta}, d_i)$ is bounded in probability. Plugging this back in Equation \ref{eq_trianineq} we consequently have that 
$$\Big|Q_n(s(\bm{\theta})) - Q_0(s(\bm{\theta}))\Big| = \mathcal{O}_p\left(\max\left( \frac{1}{k}, \frac{1}{\sqrt{n}}\right)\right),$$
which concludes the proof.
\end{proof}

\subsection{Proof of Corollary \ref{corol_sym}}
\label{app.cor}

\begin{proof}
This corollary is simply a consequence of the proof of Proposition 1. Indeed, if $s(\bm{\theta}, d_i)$ follows a symmetric distribution, we have that
$$\frac{1}{n}\sum_{i=1}^n s_i(\bm{\theta})^3 = \mathcal{O}_p\left(\frac{1}{\sqrt{n}}\right) ,$$
hence, following (\ref{eq_trianineq}) we would have
$$\Big|Q_n(s(\bm{\theta})) - Q_0(s(\bm{\theta}))\Big| \leq \mathcal{O}_p\left(\frac{1}{\sqrt{n}}\right) + \mathcal{O}_p\left(\frac{1}{k\sqrt{n}}\right) = \mathcal{O}_p\left(\text{max}\left(1, \frac{1}{k}\right)\frac{1}{\sqrt{n}}\right) .$$
In order for this term to go to zero as $n \to \infty$, we need either of the following cases:
\begin{enumerate}
	\item $k \to \infty$ (or in any case $k \geq 1$),
	\item $k \to 0$ slower than $\sqrt{n}$.
\end{enumerate}

\end{proof}

\end{document}